\documentclass[twocolumn,showpacs,preprintnumbers,amsmath,amssymb]{revtex4}

\usepackage{graphicx}
\begin{document}

%\preprint{}

\title{ Weak coupling BCS suerconductivity in the Electron-Doped Cuprate Superconductors}

\author{L. Shan$^{1}$}
  \email{shanlei@ssc.iphy.ac.cn}
\author{Y. Huang$^1$, Y. L. Wang$^1$, S. L. Li$^2$, J. Zhao$^2$, Pengcheng Dai$^{2,3}$, Y. Z. Zhang$^1$, C. Ren$^1$}
\author{H. H. Wen$^{1}$}
  \email{hhwen@aphy.iphy.ac.cn}

\affiliation{$^1$National Laboratory for Superconductivity, Institute of Physics $\&$ Beijing
National Laboratory for Condensed Matter Physics, Chinese Academy of Sciences, P.O. Box 603,
Beijing 100080, China}

\affiliation{$^2$Department of Physics and Astronomy, The University of Tennessee, Knoxville,
Tennessee 37996-1200, USA}

\affiliation{$^3$Neutron Scattering Sciences Division, Oak Ridge National Laboratory, Oak Ridge,
Tennessee
37831-6393, USA }

\date{\today}

\begin{abstract}
We use in-plane tunneling spectroscopy to study the temperature dependence of the local
superconducting gap $\Delta(T)$ in electron-doped copper oxides with various $T_c$'s and Ce-doping concentrations.
We show that the temperature dependence of $\Delta(T)$ follows the expectation of the Bardeen-Cooper-Schrieffer (BCS)
theory of superconductivity, where $\Delta(0)/k_{B}T_{c}\approx1.72\pm 0.15$ and $\Delta(0)$ is the
average superconducting gap across the Fermi surface,
 for all the doping levels investigated.
 These results suggest that the electron-doped superconducting copper oxides
are weak coupling BCS superconductors.
\end{abstract}

\pacs{74.50.+r, 74.72.Jt, 74.45.+c}

\maketitle


\section{Introduction}

The physics of conventional superconductors can be well understood by the Bardeen-Cooper-Schrieffer (BCS)
theory of superconductivity.  Within the BCS model, the
superconducting gap $\Delta(T)$ is weakly temperature dependent at low temperatures but closes
rapidly to zero near $T_{c}$. In the weak coupling limit, $\Delta(0)/k_{B}T_{c}$ is 1.76 for an
isotropic gap and becomes slightly smaller for an anisotropic gap, where $\Delta(0)$ is the zero-temperature
gap averaged over the entire Fermi surface. Since there is no generally accepted microscopic theory for
high-transition temperature (high-$T_{c}$) copper oxides, it would be interesting to see if the BCS theory can
under certain conditions describe the physics of some high-$T_c$ cuprates.
For hole-doped ($p$-type) materials, a pseudogap appears at the antinodal region
and may compete with the superconducting gap on the Fermi surface
\cite{Damascelli2003,Deutscher2005}.
Electron-doped ($n$-type) copper oxides, on the other hand,
 have a much weaker pseudogap effect and thus provide a good opportunity to investigate the superconducting gap without the influence of
``Fermi arc", ``Nodal metal" or ``Pseudogap"
\cite{Damascelli2003,Deutscher2005,Norman2005,Wen2005,Kanigel2006}.
Using angle resolved photoemission spectroscopy and transport measurements, previous work have found that
the Fermi surfaces in the $n$-type cuprates have two-band characteristics
\cite{Armitage2002,Yuan2004,Luo2005,Liu2006,Shan2005} and the superconducting gap has
the non-monotonic $d$-wave pairing
symmetry \cite{Tsuei2000,Blumberg2002,Matsui2005b}. Unfortunately, the superconducting gap of
$n$-type cuprates is relatively small and determination of its exact value over a large doping
range is an experimental challenge \cite{Kleefisch2001,Kashiyawa1998,Huang1990,Biswas2002,Blumberg2002,Matsui2005b}.
Previous estimates suggest that the superconducting pairing strength of the
$n$-type cuprates is close to a weak-coupling regime in the optimally doped and overdoped region
\cite{ZimmersA2004,QazilbashMM2005}. Furthermore, tunneling data suggest that the superconducting gap increases monotonically with decreasing doping levels even
in the underdoped regime \cite{Biswas2002}. Although much is known about these electron-doped cuprates,
there have been no systematic study of the superconducting gap as a functon of electron-doping.

In this paper,we present the point contact spectra measured on $n$-type cuprates
$Nd_{2-x}Ce_{x}CuO_{4-y}$ (NCCO) and $Pr_{1-x}LaCe_{x}CuO_{4-y}$ (PLCCO) over a wide doping range.
We find that the temperature dependence of the average gap follows the BCS predictions
with a universal weak coupling ratio
for all doping levels investigated.

\section{experimental}

\begin{table}
\caption{\label{tab:table1} Main superconducting phases studied in this work.}
\begin{ruledtabular}
\begin{tabular}{cccc}
Label  & Annealing & $T_{c}$(K) & Formula\\
\hline
          plcco-un21   & underdoped & 21 & $Pr_{0.88}LaCe_{0.12}CuO_{4-y}$\\
          plcco-un24   & underdoped & 24 &$Pr_{0.88}LaCe_{0.12}CuO_{4-y}$ \\
           ncco-op25   & optimally doped & 25 & $Nd_{1.85}Ce_{0.15}CuO_{4-y}$\\
          plcco-ov17   & overdoped & 17 & $Pr_{0.85}LaCe_{0.15}CuO_{4-y}$\\
          plcco-ov13   & overdoped & 13 & $Pr_{0.85}LaCe_{0.15}CuO_{4-y}$\\
\end{tabular}
\end{ruledtabular}
\end{table}

High-quality single crystals of NCCO and PLCCO were grown by the traveling-solvent floating-zone
technique. As grown, the crystals are not superconducting. By annealing the samples at different
temperatures in pure Ar or vacuum, bulk superconductivity with different $T_{c}$'s can be obtained
\cite{Kang2007}. The detailed information of the superconducting phases studied in this work are
presented in Table~\ref{tab:table1}. The in-plane point contact junctions were made by approaching
the Pt/ Ir alloy or Au tips towards the (100) and/or (110) surfaces of the single-crystal samples
(As discussed in this paper, there is no obvious difference between these two directions). The
tip's preparation and the details of the experimental setup were described elsewhere
\cite{Shan2003}. In order to obtain high-quality junctions with good reproducibility, the samples
were carefully processed by nonaqueous chemical etching before being mounted on the point contact
device \cite{Shan2005}. For each superconducting phase, we repeated measurements many times at
different locations on the sample surface and obtained the local superconducting gap $\Delta$ and
transition temperature $T_{c}$ which vary slightly around the bulk values.

\section{results and discussions}

\begin{figure}[top]
\includegraphics[scale=1.5]{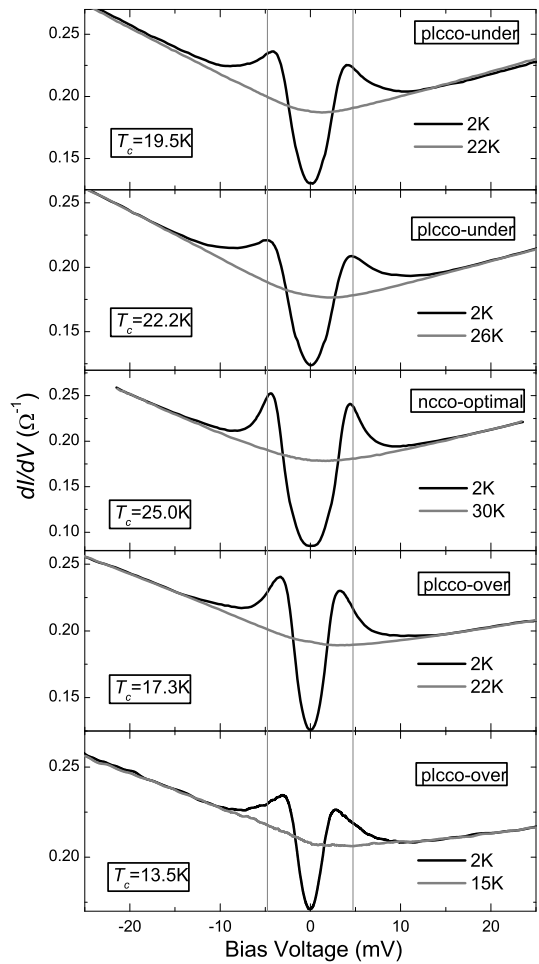}
\caption{\label{fig:fig1} Point-contact spectra measured at 2K and well above $T_{c}$ for various
doping levels. The conductance drop has been subtracted from the spectra above $T_{c}$ as
explained in the text. Those two vertical grey lines are added as references to identify the
positions of the coherence peaks. }
\end{figure}

Fig.~\ref{fig:fig1} shows the point contact spectra measured at various doping levels.
These spectra are highly reproducible on the sample surface. In each run of the
measurement, the spectra were recorded at various temperatures between 2 K and $T_{c}$ with
increments of 1 K (refer to Fig.~\ref{fig:fig3}(a)-(c)). The data measured at $T=2$ K and well
above $T_{c}$ are presented in Fig.~\ref{fig:fig1} for clarity. On the low temperature spectra, two
coherence peaks are accompanied by low-energy depression of the quasiparticle
density of states. The conductance within the gap voltage does not go to zero because the junctions
are not tunnel junctions but ballistic point contact junctions with a finite tunneling barrier
\cite{Blonder1982,Tanaka1995}. One puzzling aspect of the data is that no any zero bias conductance
peak (ZBCP) is found along the nodal direction for all doping levels, inconsistent
with the expectation of
 non-monotonic $d$-wave pairing symmetry for $n$-type cuprates
\cite{Tsuei2000,Blumberg2002,Matsui2005b,Tanaka1995}. Instead, the spectra
show almost identical shape with two distinct coherence peaks. This is difficult to understand within the current $d$-wave theory. This may be caused by the micro roughness (or corrugation) of the
nodal surface arising from the inherent crystal-lattice structure and the axis dependent
strength of chemical bonding. In this case, the dominated incident current is actually injected
along antinodal direction of the sample, and hence the identical spectral shape for both nodal and
antinodal directions can be easily understood.

It is well known that when temperature rises across the superconducting transition, the junction
becomes normal and the background conductance drops notably due to a finite normal-state resistance
of the sample ($R_s$) in series with the junction \cite{Alff2003}. In this work, $R_s$ varies from
0 to 2 ohm depending on the samples' property and the configuration of the electrodes and the point
contacts. It can be subtracted simply by replacing $dV/dI$ and $V$ by $dV/dI-R_s$ and $V-IR_s$,
respectively. All spectra (above $T_{c}$) shown in Fig.~\ref{fig:fig1} have been treated in this
way. Accordingly, we can readily construct a universal background and normalize the spectra below
$T_{c}$ which were then compared with theoretical models. In addition to
the rough estimation of $2\Delta$ from the peak-to-peak distance of the spectra (it is well known
that such method always overestimates the gap value when the measured spectrum is not ideal, in
this work such overestimate can exceed 1 meV), we used the BTK theory \cite{Blonder1982}
to derive $\Delta$ more accurately  \cite{Matsui2003,Hoffman2002}. In this model, two parameters are
introduced to describe the effective potential barrier ($Z$) and the superconducting energy gap
($\Delta$). As a supplement, the quasiparticle energy $E$ is replaced by $E+i\Gamma$, where
$\Gamma$ is the broadening parameter characterizing the finite lifetime of the quasiparticles
\cite{DynesRC1984,PlecenA1994}. For the extended anisotropic BTK model \cite{Tanaka1995}, another
parameter of $\alpha$ was introduced to distinguish different tunneling directions. In this work,
$\alpha$ is set to 0 corresponding to the antinodal direction as discussed above.

\begin{figure}
\includegraphics[scale=1.1]{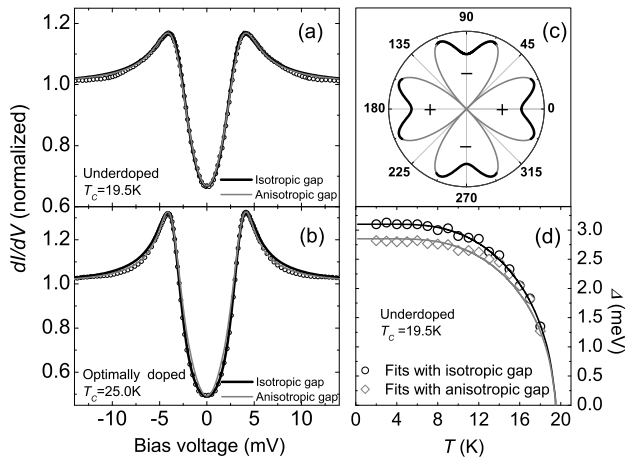}
\caption{\label{fig:fig2} Comparison of experimental data and theoretical calculations with the
isotropic BTK model (black lines) and the anisotropic one (grey lines) for two samples: (a)
plcco-un21, (b) ncco-op25. (c) The schematic diagram of the anisotropic gap (or non-monotonic
$d$-wave gap), in which the black fragments indicate the antinodal region and the grey fragments
indicate the nodal region. (d) $\Delta$(averaged gap)$\sim T$ relations determined with two
different models, solid lines are guides to eyes. }
\end{figure}

\begin{figure}
\includegraphics[scale=1.0]{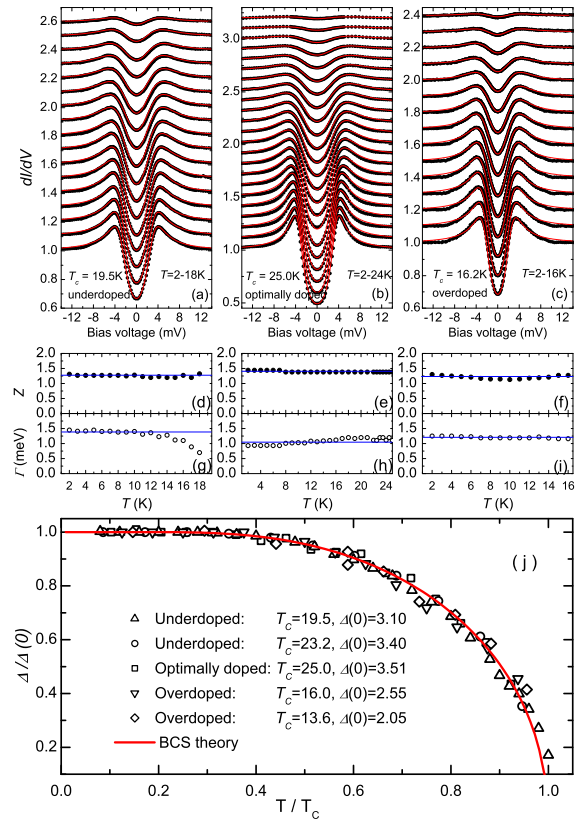}
\caption{\label{fig:fig3} (Color online) Temperature dependence of the normalized spectra (balck
dots) and theoretical calculations with the isotropic BTK model (solid lines) for: (a) plcco-un21,
(b) ncco-op25, and (c) plcco-ov17, all curves except the lowest one are shifted upwards for
clarity; (d)-(f) Fitting parameter of $Z$ corresponding to the data shown in (a)-(c) respectively;
(g)-(i) Fitting parameter of $\Gamma$ corresponding to the data shown in (a)-(c) respectively; (j)
The temperature dependence of the superconducting gap in a reduced scale for various doping
levels. }
\end{figure}

In Fig.~\ref{fig:fig2}(a) and (b), we present the comparison between the experimental data and the
theoretical calculations for both isotopic and anisotropic BTK models.
Fig.~\ref{fig:fig2}(c) illustrates the gap function of the non-monotonic $d$-wave pairing symmetry
expressed explicitly by $\Delta_{anis}= \Delta_{0}[1.43cos(2\phi)-0.43cos(6\phi)]$
\cite{Blumberg2002,Matsui2005b}. In some cases, the nodal region (the grey segments) has negligible
contribution compared to the antinodal region (the black segments) because of the weaker spectral
weight \cite{Armitage2002} or fewer charge carriers \cite{Luo2005,Shan2005} possessed by the nodal
Fermi pockets. As the first order approximation, we only consider the contribution from the antinodal
region and describe the weakly angle dependent gap as a constant value, which is equal to the
isotropic BTK model. As shown in Fig.~\ref{fig:fig2}(a), both the isotropic BTK model
and the non-monotonic $d$-wave one fit the data very well, possibly due to the
broadening effect (a finite $\Gamma$ value). If we look at the spectra shown in
Fig.~\ref{fig:fig2}(b) which has a smaller broadening effect ($\Gamma/\Delta \approx 0.28$)
than that in Fig.~\ref{fig:fig2}(a) ($\Gamma/\Delta \approx 0.44$),  the
non-monotonic $d$-wave model fits the data better at higher energy outside the coherence peaks
while the isotropic model is more favorable at lower energy. This is consistent with the
above discussions, i.e., the contribution from the antinodal Fermi pockets is dominant in this
spectrum and the constant gap is a good approximation. In
Fig.~\ref{fig:fig2}(d), we present the temperature dependence of the averaged superconducting gap determined by
both models. We find that the averaged gap determined by the isotropic model has a very
small uncertainty below $10\%$ for all the doping levels.

Fig.~\ref{fig:fig3} illustrates the details of our data analysis. Fig.~\ref{fig:fig3}(a)-(c) show
the temperature dependence of the normalized spectra of an underdoped sample, the optitimally doped
one and an overdoped one, respectively.  These data are consistent with the calculations based on
the isotropic BTK model (denoted by the solid lines). As shown in Fig.~\ref{fig:fig3}(d)-(i), the
value of the fitted barrier strength ($Z$) lies between $1.0\sim 1.5$ and $\Gamma$ value changes in
a range from 0.7 meV to 1.5 meV without an obvious dependence on the doping levels. The
independence of $Z$ on temperature indicates high stability of the junctions. For the optimally
doped and overdoped samples, $\Gamma$ is also almost independent on temperature and increases with
junction resistance similar to that of conventional BCS superconductors such as Nb \cite{shan2006}
and Zn \cite{Naidyuk1996}. However, the $\Gamma$-value of the underdoped sample decreases
continuously when temperature increases very close to $T_c$. The temperature dependence of $\Gamma$
may be closely related to the inhomogeneity, impurities, disorders, and scattering mechanism in
this system, which need to be clarified by future experiments. In fact, it has been demonstrated
that the $\Gamma$ value is not an obstacle to derive the gap value for a wide temperature scope
\cite{Shan2005}. For example, when $\Gamma$ just exceeds $\Delta$ ($\Gamma/\Delta\approx 1.1$), the
uncertainty of $\Delta$ is still below $30\%$. However, there is another factor making the fitting
procedure more difficult, which was called as ``critical current effect" \cite{Sheet2004}, because
it will distort the shape of the spectra around $T_c$. Therefore, the upper temperature limit of
our analysis is mainly determined by such effect. Accordingly, we gave up to fit the spectra for
the temperatures very close to $T_c$, as presented in Fig.~\ref{fig:fig3}(a)-(c).

\begin{figure}
\includegraphics[scale=1.2]{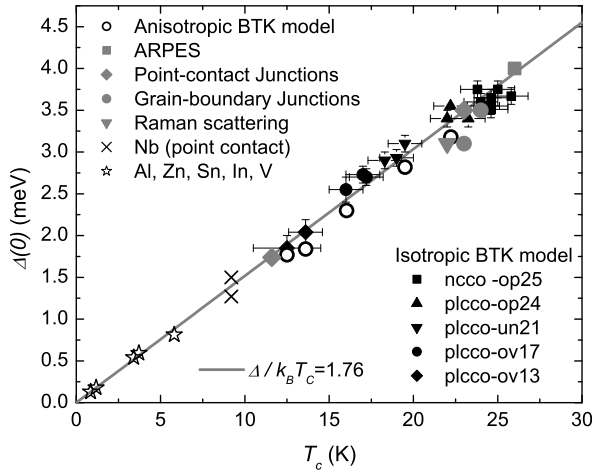}
\caption{\label{fig:fig4} Relationship between the zero-temperature averaged gap and transition
temperature. The grey symbols indicate previous reports from ARPES \cite{Matsui2005b}, Raman
\cite{Blumberg2002}, point contact \cite{Huang1990,Biswas2002} and grain-boundary
\cite{Kleefisch2001}, respectively. The open circles denote the results of fitting to the
non-monotonic $d$-wave model. The crosses indicate the data of Nb measured by the same
experimental apparatus used in this work. The data of some conventional superconductors are also
presented as open stars for comparison \cite{Wolf1985}. }
\end{figure}

It is generally accepted that a practical point contact junction often includes many channels or
real point contacts, so it is difficult to estimate the contact area simply according to the
Sharvin formula. Alternatively, we have taken a prudential way to keep our measurements away from
artificial errors. As elaborated in previous reference \cite{Blonder1983}, after the metal tip
reaches the sample surface, the barrier layer is abraded at first and its thickness decreases
slightly with the increasing pressure. Consequently, the measured spectrum becomes sharper due to
the weakening of quasiparticle scattering near the normal-metal/superconductor micro-constriction
and the barrier strength also decreases. Further pressure of the tip on the sample surface may
simply flatten the point (validating more real point contact channels) over the same minimal
thickness of a tenacious barrier layer. In this case, if no any artificial effect shows up, there
should be no obvious change on the measured spectrum in a range of junction resistance (i.e., the
normalized spectra should be identical) until the junction is damaged eventually. All measurements
are carefully checked to be in such a regime. We have also found that the determined gap value is
independent on the barrier strength and the measured locations (taking into account the slight
variation of local $T_c$), indicating that we stood a good chance to detect the bulk properties of
the samples.

The temperature dependence of superconducting gap are presented in Fig.~\ref{fig:fig3}(j) in a
reduced scale. The universal $\Delta(T)$ relation was found for all studied
doping levels in good agreement with the BCS theory. The derived $\Delta(0)$ and $T_{c}$ for
various locations on different samples are summarized in Fig.~\ref{fig:fig4}. For comparison, we
also replot the data in previous reports (grey symbols)
\cite{Kleefisch2001,Huang1990,Biswas2002,Blumberg2002,Matsui2005b} for optimally doped samples and
an overdoped one. The coupling ratio is almost a constant with a value around 1.7. If the
non-monotonic $d$-wave model is used, this ratio becomes about 1.6 (as exemplified by the open
circles). Both cases belong to the weak coupling regime. As an example to demonstrate the validity
of the methodology for determining the superconducting gap, we also presented in
Fig.~\ref{fig:fig4} the data from Nb-tip/Au-foil point contacts using the same experimental
apparatus. The determined coupling ratio is consistent with the previous reports \cite{Wolf1985}
and BCS theory with a high precision.

\section{summary}

In summary, by investigating the point contact spectra of the electron-doped cuprates, we show that
the temperature dependence of the superconducting gap follows the BCS prediction very well in a
wide doping regime with a universal weak coupling ratio of $\Delta(0)/k_{B}T_{c}= 1.72\pm 0.15$.
Therefore, the electron-doped cuprates are weak coupling BCS superconductors although the
non-monotonic $d$-wave pairing symmetry may be favorable.

Note added: During the preparation of this manuscript, we became aware that a recent report
indicates the weak coupling BCS dirty superconductivity in an electron doped cuprate based on the
measurements of SIS' junctions\cite{Dagan2007}. We also became aware of the recent STM paper (to be
published) reporting larger gaps than that in this work.

\begin{acknowledgments}
This work is supported by the National Science Foundation of China, the Ministry of Science and
Technology of China ( 973 project No: 2006CB601000, 2006CB921802, 2006CB921300 ), and Chinese
Academy of Sciences (Project ITSNEM). The PLCCO and NCCO single-crystal growth at UT is supported
by the US DOE BES under contract No. DE-FG-02-05ER46202.
We acknowledge the fruitful discussions
with Zhi-Xun Shen and N. Nagaosa.
\end{acknowledgments}

%\noindent Email address: $^*$shanlei@ssc.iphy.ac.cn\\ Email address: $^\dag$hhwen@aphy.iphy.ac.cn
%\bibliography{NoEndingPoint}

\begin{thebibliography}{}


\bibitem{Damascelli2003} A. Damascelli, Z. Hussain, and Z.X. Shen, Rev. Mod. Phys. {\bf 75},
473(2005) and references therein.
\bibitem{Deutscher2005} G. Deutscher, Rev. Mod. Phys. \textbf{77}, 109 (2005).
\bibitem{Wen2005} Hai-Hu Wen, Lei Shan, Xiao-Gang Wen, Yue Wang, Hong Gao, Zhi-Yong Liu, Fang Zhou, Jiwu Xiong, and
Wenxin Ti, Phys. Rev. B \textbf{72}, 134507 (2005).
\bibitem{Norman2005} M. R. Norman, D. Pines and C. Kallin, Advances in Physics, {\bf 54}, 715
(2005).
\bibitem{Kanigel2006}  A. Kanigel, M. R. Norman, M. Randeria, U. Chatterjee, S. Suoma, A. Kaminski, H. M. Fretwell, S. Rosenkranz, M. Shi, T. Sato, T. Takahashi, Z. Z. Li, H. Raffy, K. Kadowaki, D. Hinks, L. Ozyuzer, J. C. Campuzano
, Nature Physics {\bf 2}, 447 (2006).

\bibitem{Armitage2002} N. P. Armitage, F. Ronning, D. H. Lu, C. Kim, A. Damascelli, K. M. Shen, D. L. Feng, H. Eisaki, Z.-X. Shen, P. K. Mang, N. Kaneko, M. Greven, Y. Onose, Y. Taguchi, and Y. Tokura, Phys. Rev. Lett. {\bf 88}, 257001 (2002).

\bibitem{Yuan2004} Qingshan Yuan, Yan Chen, T. K. Lee, and C. S. Ting, Phys. Rev. B \textbf{69}, 214523 (2004).

\bibitem{Luo2005} H. G. Luo and T. Xiang, Phys. Rev. Lett. \textbf{94}, 027001 (2005).
\bibitem{Liu2006} C. S. Liu, H. G. Luo, W. C. Wu, and T. Xiang, Phys. Rev. B \textbf{73}, 174517 (2006).
\bibitem{Shan2005} L. Shan, Y. Huang, H. Gao, Y. Wang, S. L. Li, P. C. Dai, F. Zhou, J. W. Xiong, W. X. Ti, and H. H. Wen, Phys. Rev. B \textbf{72}, 144506 (2005).


\bibitem{Tsuei2000} C. C. Tsuei and J. R. Kirtley, Phys. Rev. Lett. \textbf{85}, 182 (2000).
\bibitem{Blumberg2002} G. Blumberg, A. Koitzsch, A. Gozar, B. S. Dennis, C. A. Kendziora, P. Fournier, and R. L. Greene, Phys. Rev. Lett. \textbf{88}, 107002 (2002).
\bibitem{Matsui2005b} H. Matsui, K. Terashima, T. Sato, T. Takahashi, M. Fujita, and K. Yamada, Phys. Rev. Lett. \textbf{95}, 017003 (2005).

\bibitem{Kleefisch2001} S. Kleefisch, B. Welter, A. Marx, L. Alff, R. Gross, M. Naito, Phys. Rev. B \textbf{63}, 100507(R) (2001).
\bibitem{Kashiyawa1998} S. Kashiwaya, T. Ito, K. Oka, S. Ueno, H. Takashima, M. Koyanagi, Y.
Tanaka, K. Kajimura, Phys. Rev. B \textbf{57}, 8680 (1998).

\bibitem{Huang1990} Q. Huang, J. F. Zasadzinski, N. Tralshawala, K. E. Gray, D. G. Hinks, J. L. Peng, R. L. Greene, Nature {\bf 347}, 369 (1990).
\bibitem{Biswas2002} Amlan Biswas, P. Fournier, M. M. Qazilbash, V. N. Smolyaninova, Hamza Balci, and R. L. Greene, Phys. Rev. Lett. \textbf{88}, 207004 (2002).


\bibitem{ZimmersA2004} A. Zimmers, R. P. S. M. Lobo, N. Bontemps, C. C. Homes, M. C. Barr, Y. Dagan, and R. L. Greene, Phys. Rev. B \textbf{70}, 132502 (2004).
\bibitem{QazilbashMM2005} M. M. Qazilbash, A. Koitzsch, B. S. Dennis, A. Gozar, Hamza Balci, C. A. Kendziora, R. L. Greene, and G. Blumberg, Phys. Rev. B \textbf{72}, 214510 (2005).

\bibitem{Kang2007} Hye Jung Kang, Pengcheng Dai, Branton J. Campbell, Peter J. Chupas, Stephan Rosenkranz, Peter L. Lee, Qingzhen Huang, Shiliang Li, Seiki Komiya, Yoichi Ando, Nature Mater. 6, 224 (2007) and references therein.
\bibitem{Shan2003} L. Shan, H. J. Tao, H. Gao, Z. Z. Li, Z. A. Ren, G. C. Che, and H. H. Wen, Phys. Rev. B {\bf 68}, 144510 (2003).
\bibitem{Blonder1982} G.E. Blonder, M. Tinkham, and T.M. Klapwijk, Phys. Rev. B \textbf{25}, 4515 (1982).
\bibitem{Tanaka1995} Y. Tanaka and S. Kashiwaya, Phys. Rev. Lett. \textbf{74}, 3451
(1995); Phys. Rev. B \textbf{53}, 9371 (1996).

\bibitem{Alff2003} L. Alff, Y. Krockenberger, B. Welter, M. Schonecke, R. Gross, D. Manske, M. Naito, Nature {\bf 422}, 698 (2003).

\bibitem{Matsui2003} H. Matsui, T. Sato, T. Takahashi, S.-C.Wang, H.-B. Yang, H. Ding, T. Fujii, T.Watanabe, and A. Matsuda, Phys. Rev. Lett. {\bf 90}, 217002 (2003).
\bibitem{Hoffman2002} J. E. Hoffman, K. McElroy, D.-H. Lee, K. M Lang, H. Eisaki, S. Uchida, and J. C. Davis, Science {\bf 297}, 1148 (2002).
\bibitem{DynesRC1984} R. C. Dynes, J. P. Garno, G. B. Hertel, T. P. Orlando, Phys. Rev. Lett. \textbf{53}, 2437 (1984).
\bibitem{PlecenA1994} A. Plecen\'{i}k, M. Grajcar, \v{S}. Be\v{n}a\v{c}ka, P. Seidel and A. Pfuch, Phys. Rev. B \textbf{49}, 10016 (1994).
\bibitem{Wolf1985} E. L. Wolf, {\it Principles of Electron Tunneling Spectroscopy.} (Oxford Univ. Press, New York,
1985).

\bibitem{shan2006} L. Shan, Y. Huang, C. Ren, and H. H. Wen, Phys. Rev. B {\bf 73}, 134508 (2006).
\bibitem{Naidyuk1996} Yu. G. Naidyuk, H. v. L\"{o}hneysen, I. K. Yanson, Phys. Rev. B {\bf 54}, 16077 (1996).


\bibitem{Sheet2004} G. Sheet, S. Mukhopadhyay, and P. Raychaudhuri, Phys. Rev. B {\bf 69}, 134507 (2004).

\bibitem{Blonder1983} G. E. Blonder and M. Tinkham, Phys. Rev. B {\bf 27}, 112 (1983).

\bibitem{Dagan2007}Y. Dagan, R. Bek, R. L. Greene, Condmat/0702091.



\end{thebibliography}

\end{document}